\documentclass[a4paper,11pt]{article}
\usepackage[top=3cm,bottom=3.4cm,left=2cm,right=2cm]{geometry}

\usepackage{color}
\usepackage{bm}
\usepackage{cite}
\usepackage{amsmath}
\usepackage{amssymb}
\usepackage{cite}
\usepackage[dvipdfmx]{graphicx}
\DeclareFontFamily{OT1}{pzc}{}
\DeclareFontShape{OT1}{pzc}{m}{it}{<-> s * [1.10] pzcmi7t}{}
\DeclareMathAlphabet{\mathpzc}{OT1}{pzc}{m}{it}
\newcommand{\nn}{\nonumber\\}
\newcommand{\bra}[1]{\!\left<#1 \right|}
\newcommand{\ket}[1]{\left| #1 \right>\!}

\renewcommand{\thepage}{}
\makeatletter
\@addtoreset{equation}{section}
\renewcommand{\theequation}{\thesection.\@arabic\c@equation}
\makeatother
\renewcommand{\thefootnote}{\fnsymbol{footnote}}

\begin{document}
\begin{titlepage}
\title{
\vspace*{-4ex}
\hfill
\begin{minipage}{3.5cm}
\end{minipage}\\
 \bf 
Two-point closed  string amplitudes\\
 in the BRST formalism
\vspace{0.5em}
}

\author{
Isao~{\sc Kishimoto},$^{1}$\footnote{\tt ikishimo@rs.socu.ac.jp}
~~~Shigenori {\sc Seki},$^{2}$\footnote{\tt sigenori@yukawa.kyoto-u.ac.jp}
~~~Tomohiko~{\sc Takahashi}$^{3}$\footnote{\tt tomo@asuka.phys.nara-wu.ac.jp}
\\
\vspace{0ex}\\
\\
$^{1}${\it
 Center for Liberal Arts and Sciences, Sanyo-Onoda City University,}\\
{\it Daigakudori 1-1-1, Sanyo-Onoda Yamaguchi 756-0884, Japan}\\
\vspace{1ex}
\\
$^{2}${\it Osaka Central Advanced Mathematical Institute (OCAMI),}\\
{\it Osaka Metropolitan University,}\\
{\it 3-3-138, Sugimoto, Sumiyoshi-ku, Osaka 558-8585, Japan}\\
\vspace{1ex}
\\
$^{3}${\it Department of Physics, Nara Women's University,}\\
{\it Nara 630-8506, Japan}
\vspace{0ex}
}

\date{}
\maketitle

\begin{abstract}
\normalsize

Two-point tree-level amplitudes in bosonic closed string theory are described by a correlation function within the BRST formalism, which respects manifest Lorentz and conformal invariance.
In the derivation of the two-point amplitudes, we use the mostly BRST exact operator, which has been introduced for two-point open string amplitudes, and a closed string vertex with ghost number three, which has been explored in our recent work,  in addition to the conventional one.

\end{abstract}

\end{titlepage}

\renewcommand{\thepage}{\arabic{page}}
\renewcommand{\thefootnote}{\arabic{footnote}}
\setcounter{page}{1}
\setcounter{footnote}{0}


\section{Introduction}

In string theory, the scattering matrix element is computed by employing
the Polyakov path integral expression.  In \cite{Erbin:2019uiz},
it is demonstrated that
even the two-point open string amplitude can be computed using the same
approach, despite the presence of an infinite volume for the residual
symmetry after fixing two positions. The resulting amplitude agrees with
standard free particle expression in quantum field theory.

In \cite{Seki:2019ycz}, it is shown that the two-point open string
amplitude is calculated also in the BRST operator formalism.  When the
positions of two open strings are fixed in the upper half-plane, the
vertex operators have the ghost number two, which is insufficient to
yield non-zero amplitudes.  Then, in \cite{Seki:2019ycz}, a mostly BRST
exact operator is introduced to fix the residual symmetry and add
ghost number one.  The two-point open string amplitude is provided by
calculating the three-point function in which two on-shell open string
vertex operators and the supplementary operator are inserted in the
worldsheet. In \cite{Kashyap:2022qtd}, 
two-point superstring amplitude is calculated in the pure spinor formalism
by using a supersymmetric mostly BRST exact operator.

The three-point open string amplitudes can be also derived from the four-point
 function with the mostly BRST exact operator \cite{Seki:2021ivm}.
Unlike the two-point amplitude, the Feynman $i\varepsilon$ prescription
has an important role in this derivation.  Along the similar
prescription, the Veneziano amplitude, namely the four-point open string
amplitude, is also calculated from the five-point function by examining
the moduli space \cite{Kishimoto:2021csf}.

Two-point closed string amplitude is also examined in the initial paper
\cite{Erbin:2019uiz}, with specific details provided in
\cite{Erbin:2021smf}.  It appears that the amplitude has been derived,
but the calculation depends on a specific coordinate frame, with two
vertices placed at the origin and infinity on the complex plane.  In
other words, the derivation lacks manifest conformal invariance.
Recently, there is an interesting study of considering a timelike
Liouville direction as a regularization parameter to derive the
two-point amplitude \cite{Giribet:2023gub}. This method preserves
conformal invariance, and Poincar\'e symmetry is restored by taking a
limit, although the Liouville direction breaks the symmetry.

In the
framework of the BRST formalism, two-point closed string amplitude is also
explored using the mostly BRST exact operator with ghost number one
\cite{Seki:2021ivm}.
However, it becomes necessary to adjust the unintegrated
 closed string vertex operator
in order to achieve the required ghost number, ensuring the total ghost
number of six.  Consequently, a subtle issue arises where the resulting
vertex operator is not annihilated by $b_0-\tilde{b}_0$.

Recently, closed string vertex operators have been revisited by means of
the Faddeev-Popov procedure and descent equations in the BRST formalism
\cite{rf:KKST}.  As a result, it has been revealed that the $b_0-\tilde{b}_0$ condition,
which implies the level-matching condition,
 is not necessarily required to perform the correct
gauge fixing and calculate scattering amplitudes. Therefore, based on
the vertex operator in \cite{rf:KKST}, it should be possible to 
justify the derivation of two-point closed string amplitudes
as discussed in \cite{Seki:2021ivm}.
This is a main purpose of this paper.

In this paper, we will begin by verifying the effectiveness of the
mostly BRST exact operator inserted at the bulk by computing two-point
open string amplitudes with its use.  Subsequently, we will demonstrate
that two-point closed string amplitudes can be obtained within the BRST
formalism in a similar manner.

\section{Two-point open string amplitudes}

In \cite{Seki:2019ycz,Seki:2021ivm,Kishimoto:2021csf}, the mostly BRST exact
operator is
inserted at the boundary of the worldsheet to obtain open string
amplitudes. However, if the operator at the bulk, as discussed in
\cite{Seki:2019ycz} as the closed string case, proves to be effective,
it can also be used to derive open string amplitudes.  The operator in
\cite{Seki:2019ycz} is given by\footnote{The normalization is changed 
from \cite{Seki:2019ycz} by
following \cite{Seki:2021ivm,Kishimoto:2021csf}.}
\begin{eqnarray}
 {\mathpzc E}(z,\bar{z})&\equiv& \int^\infty_{-\infty}\frac{dq}{2\pi}
\Big\{\big(c\partial X^0 +\tilde{c}\bar{\partial}X^0\big)
e^{-iqX^0}-i\alpha'\frac{q}{4}\big(\partial c+\bar{\partial}\tilde{c}
\big)e^{-iqX^0}\Big\},
\end{eqnarray}
where $X^0(z,\bar{z})$ is the $0$-th string coordinate and $c(z)$,
$\tilde{c}(\bar{z})$ are ghost fields.  This operator inserted at $z_0$
corresponds to choosing one of the gauge fixing conditions as
\begin{eqnarray}
 X^0(z_0,\bar{z}_0)=0,
\end{eqnarray}
which is the same one as discussed in \cite{Erbin:2019uiz} even
for the open string case.

The operator can be written as the mostly BRST exact expression,
\begin{eqnarray}
 {\mathpzc E}(z,\bar{z})&\equiv& \int^\infty_{-\infty}\frac{dq}{2\pi}
\,\frac{i}{q}\bm{\delta}_{\rm B} e^{-iqX^0},
\label{eq:BRSexact}
\end{eqnarray}
where $\bm{\delta}_{\rm B}$ denotes the BRST transformation.  As discussed
in \cite{Seki:2019ycz}, this expression reflects the Lorentz and
conformal invariance of the amplitude with the insertion of $\mathpzc
E$. It is noted that explicit calculations are simplified by employing this expression.

Let us consider open strings in $26$ flat spacetime directions. The
worldsheet is given by the upper half plane and the real axis
corresponds to the open string boundary.  First, we will illustrate the
validity of the bulk operator by calculating the amplitude for two open
string tachyon as follows:
\begin{eqnarray}
{\cal A} =ig_{\rm o}^2 C_{D_2}
\bra{0}{\mathpzc E}(z_0,\bar{z}_0)\,ce^{ip_1\cdot X}(x_1)
ce^{ip_2\cdot X}(x_2)\ket{0},
\label{eq:Aop}
\end{eqnarray}
where $g_{\rm o}$ is the open string coupling constant and the
normalization constant is given by $C_{D_2}=1/(\alpha'g_{\rm o}^2)$
\cite{Polchinski:1998rq}.  The point $z_0$ is located within the upper
half plane, while $x_1$ and $x_2$ lie on the real axis.
Calculating the correlation function, (\ref{eq:Aop}) can be expressed as
\begin{eqnarray}
 {\cal A}
&=&
({p_1}^0-{p_2}^0)(2\pi)^{25}\delta^{25}(\bm{p}_1+\bm{p}_2)
\int_{-\infty}^\infty dq\,\delta(q+p_1^0+p_2^0)
\nn
&&
\times
|z_0-x_1|^{-2\alpha' q p_1^0}|z_0-x_2|^{-2\alpha' q p_2^0}
|x_1-x_2|^{-\alpha' q^2}|z_{0}-\bar{z}_0|^{-\frac{1}{2}
\alpha' q^2}.
\end{eqnarray}
Here, ${p_i}^0=\pm \sqrt{\bm{p}^2_i+m^2}$ by the on-shell condition
$p^2=m^2=-1/\alpha'$, and so the momentum conservation leads to
${p_1}^0+{p_2}^0=0$ or ${p_1}^0-{p_2}^0=0$.
Hence, we observe that, due to the factor ${p_1}^0-{p_2}^0$,
 ${\cal A}$ is nonzero only
when ${p_1}^0+{p_2}^0=0$, indicating that one tachyon corresponds to
incoming while the other corresponds to outgoing.
In the nonzero case, the resulting amplitude is given by
\begin{eqnarray}
{\cal A}=\frac{{p_1}^0}{|{p_1}^0|}\times
2{p_1}^0(2\pi)^{25}\delta^{25}(\bm{p}_1+\bm{p}_2).
\label{eq:twoopen}
\end{eqnarray}
This provides the correct amplitude of two open string tachyons, with the exception of 
the sign factor, which is inevitable when working within the BRST formalism
\cite{Seki:2019ycz}.

It is worth noting that in the correlation function (\ref{eq:Aop}), we
fix two positions of the real axis and one position of the bulk, which
seems to fix four degrees of freedom. However, this does not imply excessive
fixing condition of $PSL(2,\mathbb{R})$, which has three degrees of freedom. This
result suggests that $\mathpzc E$ merely constraints one degree of
freedom within $PSL(2,\mathbb{R})$, although it is located in the bulk.

For general open string states, we can establish
the same result by applying analogous reasoning as discussed in \cite{Seki:2019ycz}.
We consider the following correlation function,
\begin{eqnarray}
{\cal A} =ig_{\rm o}^2 C_{D_2}
\bra{0}{\mathpzc E}(z_0,\bar{z}_0)\,cV_1(x_1)\,cV_2(x_2)\ket{0},
\label{eq:Aop2}
\end{eqnarray}
where $V_i(x_i)$ is a matter vertex operator with conformal weight one.
According to the energy conservation, (\ref{eq:Aop2}) is expressed as
\begin{eqnarray}
 {\cal A}=\int_{-\infty}^\infty dq\,2\pi \delta(q+{p_1}^0+{p_2}^0)\times
\cdots,
\end{eqnarray}
where ${p_i}^0$ denotes the energy of the vertex $V_i$, and ${p_i}^0=\pm
\sqrt{\bm{p}_i^2+m_i^2}$ by the on-shell condition.  If
${p_1}^0+{p_2}^0\neq 0$, $q$ is replaced by a nonzero value after the
$q$ integration. In this case, the correlation function becomes zero,
because, for $q\neq 0$, ${\mathpzc E}$ is regarded as a BRST exact
operator through the expression (\ref{eq:BRSexact}). If
${p_1}^0+{p_2}^0=0$, $q$ can be set to zero after the $q$ integration
and so ${\cal A}$ can be rewritten as
\begin{eqnarray}
{\cal A}= ig_{\rm o}^2 C_{D_2}
\bra{0}\big(c\partial X^0 +\tilde{c}\bar{\partial}X^0\big)(z_0,\bar{z}_0)
\,cV_1(x_1)\,cV_2(x_2)\ket{0}'.
\label{eq:Aop3}
\end{eqnarray}
Here, the prime implies that the delta function of the energy
conservation and the factor $2\pi$ are excluded. This is an analogous
expression of the two-point open string amplitude in
\cite{Erbin:2019uiz}.  By referring to the calculation in
\cite{Erbin:2019uiz}, we observe that this expression coincides with
(\ref{eq:twoopen}) for general open string states.

\section{Two-point closed string amplitudes}

Let us now consider the derivation of two-point closed string amplitudes
by inserting the operator ${\mathpzc E}$.  According to the conventional
approach, the correlation function consisting of two closed strings and the
operator ${\mathpzc E}$ possesses ghost number five, which is
insufficient for deriving non-zero amplitudes. This is because the
conventional fixed vertex operator of closed strings has ghost number
two. 

Hence, we replace one of the closed string vertices with
the ghost number three operator,
\begin{eqnarray}
 \frac{1}{4\pi}(\bar{\partial}\tilde{c}-\partial c)c\tilde{c}V(z,\bar{z}),
\label{rf:gh3op}
\end{eqnarray}
where $V(z,\bar{z})$ is a matter vertex.  This vertex operator is
derived in \cite{rf:KKST} through the Faddeev-Popov procedure for gauge
fixing of $PSL(2,\mathbb{R})$.  As discussed in \cite{rf:KKST}, the state
corresponding to (\ref{rf:gh3op}) is not annihilated by $b_0-\tilde{b}_0$. 
However, this operator still yields the correct amplitude. In
particular, it is noted that the factor $1/(4\pi)$ is determined by the
Faddeev-Popov procedure.

First, we will examine a two-point closed string tachyon amplitude, which should be
obtained from the following correlation function:
\begin{eqnarray}
 {\cal A}&=&
ig_{\rm c}^2 C_{S_2}
\bra{0}{\mathpzc E}(z_0,\bar{z}_0)
\,ic\tilde{c} e^{ip_1\cdot X}(z_1,\bar{z}_1)
\,\frac{1}{4\pi}(\bar{\partial}\tilde{c}-\partial c)
c\tilde{c} e^{ip_2\cdot X}(z_2,\bar{z}_2)\ket{0},
\end{eqnarray}
where the correlation function is defined over the whole complex plane, and
the bra vacuum $\bra{0}$ is taken as the hermitian conjugate to
$\ket{0}$, as defined in \cite{rf:KKST}.\footnote{As in \cite{rf:KKST},
the vacuum is normalized as
\begin{eqnarray}
 \bra{0}c(z_1)c(z_2)c(z_3)
\tilde{c}(\bar{z}_1)\tilde{c}(\bar{z}_2)\tilde{c}(\bar{z}_3)
\ket{0}=-i|z_1-z_2|^2|z_1-z_3|^2|z_2-z_3|^2\times (2\pi)^{26}\delta^{26}(0).
\end{eqnarray}
Then, the unintegrated closed string vertex operator is defined as
\begin{eqnarray}
 ic\tilde{c}V(z,\bar{z}),
\end{eqnarray}
which naturally corresponds to the integrated vertex, $d^2 z
V(z,\bar{z})=idz\wedge d\bar{z}V(z,\bar{z})$.  } $g_{\rm c}$ denotes the
closed string coupling constant and $C_{S_2}$ is the normalization
constant given in \cite{Polchinski:1998rq}.

When computing the correlation function, ${\cal A}$ can be expressed as follows:
\begin{eqnarray}
 {\cal A}&=&
ig_{\rm c}^2 C_{S_2}  (2\pi)^{25}\delta^{25}(\bm{p}_1+\bm{p}_2)
\nn
&&
\times\frac{i\alpha'}{4\pi}
\frac{{p_2}^0-{p_1}^0}{2}
\int_{-\infty}^\infty dq \,\delta(q+{p_1}^0+{p_2}^0)\,
|z_{0}-z_{1}|^{-\alpha' q{p_1}^0}
|z_{0}-z_{2}|^{-\alpha' q{p_2}^0}
|z_{1}-z_{2}|^{-\alpha' \frac{q^2}{2}}.
\end{eqnarray}
Applying the on-shell condition and employing reasoning analogous to that in the open string case, ${\cal A}$ becomes non-zero only when ${p_1}^0+{p_2}^0=0$.
In this case, the resulting amplitude is
\begin{eqnarray}
 {\cal A}
&=&
\frac{1}{2}\Big(\frac{{p_1}^0}{|{p_1}^0|}
-\frac{{p_2}^0}{|{p_2}^0|}
\Big)\times
\frac{\alpha'}{4\pi}\,
g_{\rm c}^2 C_{S_2}\,|p_1^0| (2\pi)^{25}\delta^{25}(\bm{p}_1+\bm{p}_2).
\label{eq:twoclosed}
\end{eqnarray}
We can confirm that the numerical constant agrees with the correct one
given by the standard quantum field theory. Using $C_{S_2}=8\pi/(\alpha'
g_{\rm c}^2)$ in \cite{Polchinski:1998rq}, we find that
\begin{eqnarray}
\frac{\alpha'}{4\pi}\,g_{\rm c}^2\,C_{S_2}
=2.
\end{eqnarray}
Thus, we can derive the two-point closed string tachyon amplitude by means of $\mathpzc E$,
excluding the sign factor.

For general closed string vertices $V_i(z_i,\bar{z}_i)\ (i=1,\,2)$,
the amplitude should be given by the following correlator
\begin{eqnarray}
  {\cal A}&=&
ig_{\rm c}^2 C_{S_2}
\bra{0}{\mathpzc E}(z_0,\bar{z}_0)
\,ic\tilde{c} V_1(z_1,\bar{z}_1)
\,\frac{1}{4\pi}(\bar{\partial}\tilde{c}-\partial c)
c\tilde{c} V_2(z_2,\bar{z}_2)\ket{0}.
\label{eq:Aclosed}
\end{eqnarray}
Analogous to the open string case, this correlator becomes zero
owing to the BRST invariance if ${p_1}^0+{p_2}^0\neq 0$.
When ${p_1}^0+{p_2}^0= 0$, similar to (\ref{eq:Aop3}),
it can be rewritten as a
correlator without the delta function of energy conservation:
\begin{eqnarray}
   {\cal A}&=&
ig_{\rm c}^2 C_{S_2}
\bra{0}\left(c\partial X^0+\tilde{c}\bar{\partial}X^0
\right)(z_0,\bar{z}_0)
\,ic\tilde{c} V_1(z_1,\bar{z}_1)
\,\frac{1}{4\pi}(\bar{\partial}\tilde{c}-\partial c)
c\tilde{c} V_2(z_2,\bar{z}_2)\ket{0}'.
\end{eqnarray}
Since $\partial X^0$ is a holomorphic current, we find that
\begin{eqnarray}
&&
\bra{0}c\partial X^0(z_0)
\,ic\tilde{c} V_1(z_1,\bar{z}_1)
\,\frac{1}{4\pi}(\bar{\partial}\tilde{c}-\partial c)
c\tilde{c} V_2(z_2,\bar{z}_2)\ket{0}'
\nn
&&=
-\frac{i\alpha'}{2}{p_1}^0
\frac{z_1-z_2}{(z_0-z_1)(z_0-z_2)}\bra{0}c(z_0)
\,ic\tilde{c} V_1(z_1,\bar{z}_1)
\,\frac{1}{4\pi}(\bar{\partial}\tilde{c})
c\tilde{c} V_2(z_2,\bar{z}_2)\ket{0}'.
\label{eq:correlfunc}
\end{eqnarray}
It is noted that the factor $i\alpha'/2$ is different from the
open string two-point case, where $z_1$ and $z_2$ are on the boundary.
We can compute the remaining correlation of (\ref{eq:correlfunc}) as follows:
\begin{eqnarray}
&&-\frac{i\alpha'}{8\pi}{p_1}^0
\frac{z_1-z_2}{(z_0-z_1)(z_0-z_2)}
\times (z_0-z_1)(z_0-z_2)(z_1-z_2)(\bar{z}_1-\bar{z}_2)^2
\nn
&&
\times
(2\pi)^{25}\delta^{25}(\bm{p}_1+\bm{p}_2)\frac{1}{|z_1-z_2|^4}.
\end{eqnarray}
Here, the second factor arises from the ghost expectation value, and the
third one is attributed to the matter contribution, which is determined
by conformal symmetry. 
This computation closely follows the approach
presented in \cite{Erbin:2019uiz}, where two-point open string amplitudes are
derived using a similar method.  As a result, this correlation is
independent of the positions of the vertices.  Finally, substituting this
result and its anti-holomorphic counterpart into (\ref{eq:Aclosed}), it
becomes evident that the two-point closed string amplitudes are accurately
obtained from (\ref{eq:Aclosed}), with the exception of the sign
factor.  In other words, the resulting amplitude agrees with
(\ref{eq:twoclosed}), including the numerical factor, for general closed
string vertices.

\section{Concluding remarks
\label{sec:remarks}}

We have shown that, in the BRST formalism, two-point closed string amplitudes
can be derived from the correlation function of two closed string
vertices and the mostly BRST exact operator. In this derivation, we have
used closed string vertices with ghost number three, which are
introduced in the previous study \cite{rf:KKST}. The advantage of our
derivation is evident in its preservation of Lorentz symmetry and
conformal symmetry.

In the case of introducing $\mathpzc E$ at the boundary for open string
amplitudes, as discussed in \cite{Seki:2021ivm}, the overall sign factor
can be identified as the signed intersection number associated with the
graph $u=X^0(y)$, where $y$ represents the boundary position.  It is
worth noting that the same identification applies when inserting
$\mathpzc E$ within the bulk. In this case, the signed intersection
number is associated with the graph of $u=X^0(z,\bar{z})$, where $z$
represents the position along an arbitrary path connecting two unintegrated
vertex operators on the worldsheet.

By following our formulation, it becomes relatively straightforward to
explore the two-point amplitude of closed superstrings. Furthermore, by
the conformal invariance of our approach, we can derive general
amplitudes through the correlation function with the insertion of the
operator $\mathpzc E$. We will elaborate on the specifics of this study
in the future.

\section*{Acknowledgments}
We would like to express our gratitude to Tomomi Kitade for taking an
interest in this subject during its early stages.  This work was
supported in part by JSPS Grant-in-Aid for Scientific Research (C)
\#20K03972.  I.~K.~was supported in part by JSPS Grant-in-Aid for
Scientific Research (C) \#20K03933.  S.~S.~was supported in part by MEXT
Joint Usage/Research Center on Mathematics and Theoretical Physics at
OCAMI and by JSPS Grant-in-Aid for Scientific Research (C) \#17K05421
and \#22K036125.  T.~T.~was supported in part by JSPS Grant-in-Aid
for Scientific Research (C) \#23K03388.


\end{document}